\documentclass{article}

\usepackage[english]{babel}

\usepackage[letterpaper,top=2cm,bottom=2cm,left=3cm,right=3cm,marginparwidth=1.75cm]{geometry}

\usepackage{amsmath}
\usepackage{graphicx}
\usepackage{comment}
\usepackage[colorlinks=true, allcolors=blue]{hyperref}

\title{To Jab or Not to Jab? A Study on COVID-19 Vaccine Hesitancy in India}
\author{Shagata Mukherjee$^1$, Sayantari Ghosh$^2$, Saumik Bhattacharya$^{3,\;*}$, Sujoy Chakravarty$^{4}$}
\begin{document}
\maketitle
\date{}
\noindent $^1$ Centre for Social and Behaviour Change, Ashoka University, Rajiv Gandhi Education City, Sonipat, Haryana, India.\\
$^2$ National Institute of Technology Durgapur, M.G. Avenue, Durgapur, West Bengal, India.\\
$^3$ Indian Institute of Technology Kharagpur, Kharagpur, West Bengal, India.\\
$^4$ Jawaharlal Nehru University, New Mehrauli Road, New Delhi, India.\\
$*$ Corresponding author

\begin{abstract}With a country-wide comprehensive internet survey conducted in India, we aim to determine the factors that drive hesitancy towards getting vaccinated for COVID-19, and also compare their levels of influence. The perceived reliability and effectiveness of available vaccines turn out to be important drivers in lowering vaccine hesitancy. Additionally, higher hesitancy is associated with being of a younger age or having lower education. Furthermore, comparing vaccine attitudes from observations before a major COVID-19 wave with those from after, we find that the latter are significantly less hesitant about getting the vaccine. We also find that in addition to the standard knowledge and awareness campaigns, local level peer influences are important factors that affect vaccine hesitancy. Finally, we use statistically significant estimates from logistic regression on our survey data in a synthetic heterogeneous complex network-based society, to extrapolate scenarios that may arise from the dynamic interactions between our variables of interest. We use outcomes from this simulated society to suggest strategic interventions that may lower vaccine hesitancy.
\end{abstract}
\textbf{One-Sentence Summary:} {An integrative approach with  econometric regression-guided data mining \& complex network simulations, to explore the major barriers and drivers of COVID-19 vaccination dynamics in India. }

\section{Introduction}

The COVID-19 pandemic has negatively impacted the socio-economic conditions in both developed and developing countries around the world for over two years  \cite{yeyatifilippini2021}. %\footnote{Worldometer: COVID-19 in India (As on January 9,2022) -  https://www.worldometers.info/coronavirus/country/india and https://www.worldometers.info/coronavirus/country/us/} 
With an upsurge in new cases from the Omicron variant leading to a new major wave of infections in numerous countries starting in early 2022, COVID-19 related hospitalizations and deaths are again contingencies that most governments have to be prepared for going forward \cite{callaway2021}. As the COVID-19 virus is already in community transmission, it may be a while before we see the end of waves of infection that cause social and economic disruption for some time \cite{scudellari2020, rasuletal2021}. Thus, in order to lower transmission and limit virus mutation, increasing vaccine uptake to as many people as possible has become a key governmental initiative to getting economies and societies back on track  \cite{skeggetal2021, murraypiot2021}. However, even in early 2022, only about 59\% of the total population of the world was completely or partially vaccinated. Furthermore, deployment of vaccines has not been uniform. With China and the USA having a substantial vaccination coverage (84\% and 62\% full vaccination), India (45\%) and Indonesia (42\%) are close to fully vaccinating half their populations. On the other hand, African continent as whole have managed to fully vaccinate only about 10\% of their populations.\footnote{Our World in Data (As on January 6,2022) - https://ourworldindata.org/covid-vaccinations?country=IND~USA}\\

Initially, many developing countries faced several challenges like production and supply bottlenecks, which kept vaccination rates low \cite{SousaRosa2021}. However, as both vaccine production and accessibility gradually increased, the focus on vaccine hesitancy has become important. Vaccine hesitancy has been defined by the Sage Working Group on Vaccine Hesitancy and World Health Organization (WHO) as the “delay in acceptance or refusal of vaccines despite availability of vaccine services” which cites reasons such as confidence, complacency and convenience for its existence \cite{Butler2015}. Building on this, a 5C (confidence, constraints, complacency, calculation and collective responsibility) framework  for studying the psychological antecedents of vaccine hesitancy is proposed by Betsch et al.\cite{Betsch2018}. Extant studies of vaccine hesitancy such as Sallam \cite{sallam2021covid} finds a wide variation in acceptance rates between countries. In the Indian context, vaccine hesitancy is seen to reduce over the waves of the pandemic \cite{Chandani2021, lazarus2021global} and that hesitancy significantly varies by state \cite{Soumietal2021}. Tagat et al. \cite{Tagat2022} find that mask-wearing and handwashing beliefs, information sources related to COVID-19, and past COVID-19 infection and testing status are all strongly associated with vaccination decision, while Roy Chowdhury et al. \cite{Soumietal2021} find that the main reason for hesitancy are vaccine safety, side effects, mistrust in vaccine efficiency and complacency. A more detailed literature review is given in the supplementary material. \\

As the COVID-19 pandemic continues to be an ever-changing situation, the study of vaccine hesitancy and its determinants are crucial to the management of future outbreaks. Recent research has also suggested
that as the pandemic has progressed over time, the intention
to get vaccinated has reduced \cite{robinson2021}. Our study aims to add to the existing literature by ascertaining which factors determine vaccine hesitancy in urban India \cite{guptaetal2021} through primary data collection and exhaustive data analysis using data mining and complex network simulations. We proceed to understand if COVID-19 related indicators such as an individual’s history with the disease, COVID-19 infections and deaths in one’s social circles and vaccine effectiveness perception affect their hesitancy towards vaccines. We also study the impact of behavioural factors such as time preference, risk preference and loss aversion as well as socio-demographic factors such as age, gender, location, education, occupation and family income in determining vaccine hesitancy. Our research design also makes use of the exposure of participants to a naturally occurring exogeneous disease shock: the devastating Delta wave of COVID-19, in potentially lowering vaccine hesitancy. We find that indeed participants who answer our survey after the peak of second wave infections in India \cite{ratnaetal2021} display lower hesitancy, thus highlighting a similar type of `procedural rationality' in the adoption of preventive measures for COVID-19 as is observed in citizens and law makers of Taiwan who were exposed to the SARS epidemic in the early 2000s \cite{dasguptaetal2020}. Moreover, with the help of complex network analysis, we explore the impacts of significant results from our econometric analysis on a synthetic society which is subject to a contagious COVID-19-like epidemic. Results from our simulation exercise allow us to isolate factors that can offer guidelines for better policy making in managing future COVID-19 infections as well as other disease outbreaks. Finally, by integrating our survey data with locational and contextual data that we mine, we explore whether or not individuals have a clear understanding about the real status of infection in their neighbourhoods when making decisions regarding whether or not to get vaccinated.\\

\begin{figure}
\centering
\includegraphics[width=\textwidth]{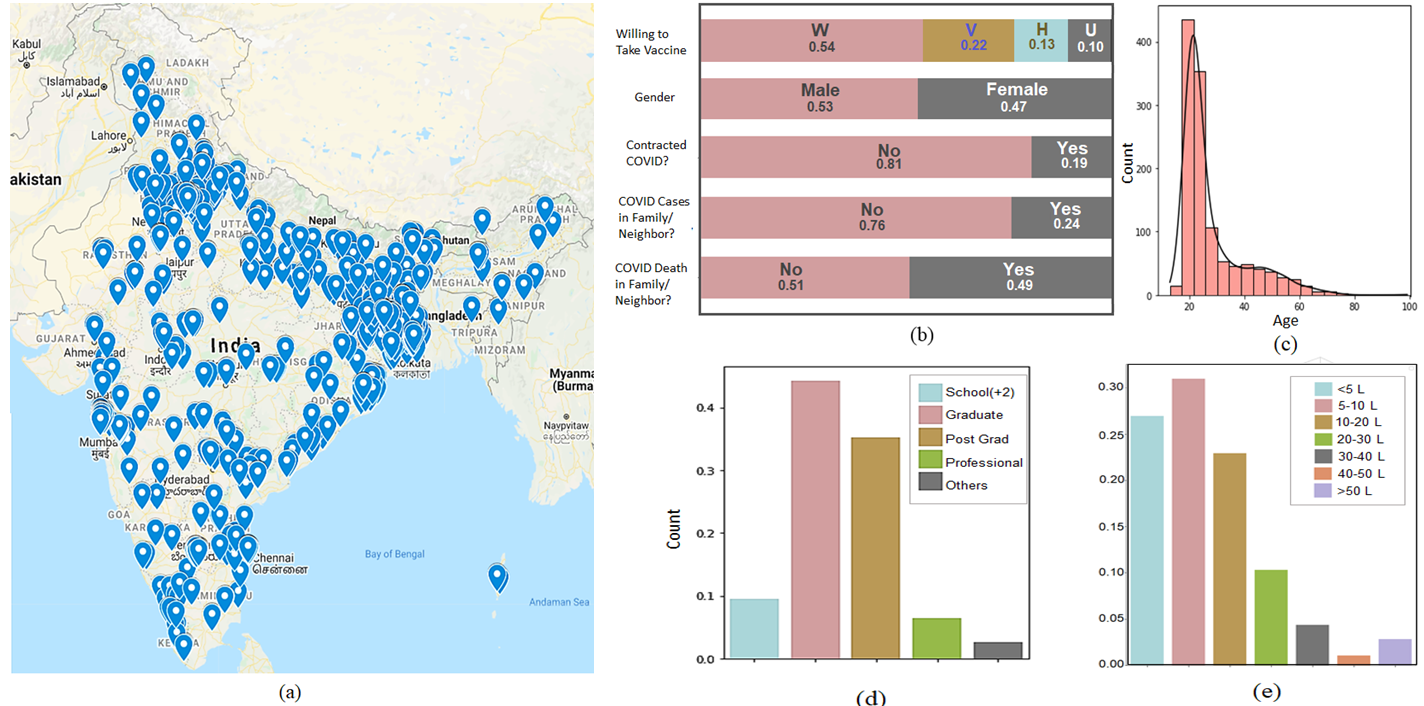}
\caption{\label{fig:fig1} General overviews of the collected data (a) The geographical locations of the respondents; (b) gender distribution, COVID-19 experiences and willingness to take the vaccine once it is available. Here, `W', `V', `H' and `U' indicate willing, vaccinated, hesitant and unwilling sub-populations; (c) Age distribution of the respondents, The black line indicates the estimated density function; (d) Academic qualification; (e) Yearly income of the family. }
\end{figure}

\begin{figure}
\centering
\includegraphics[width=\textwidth]{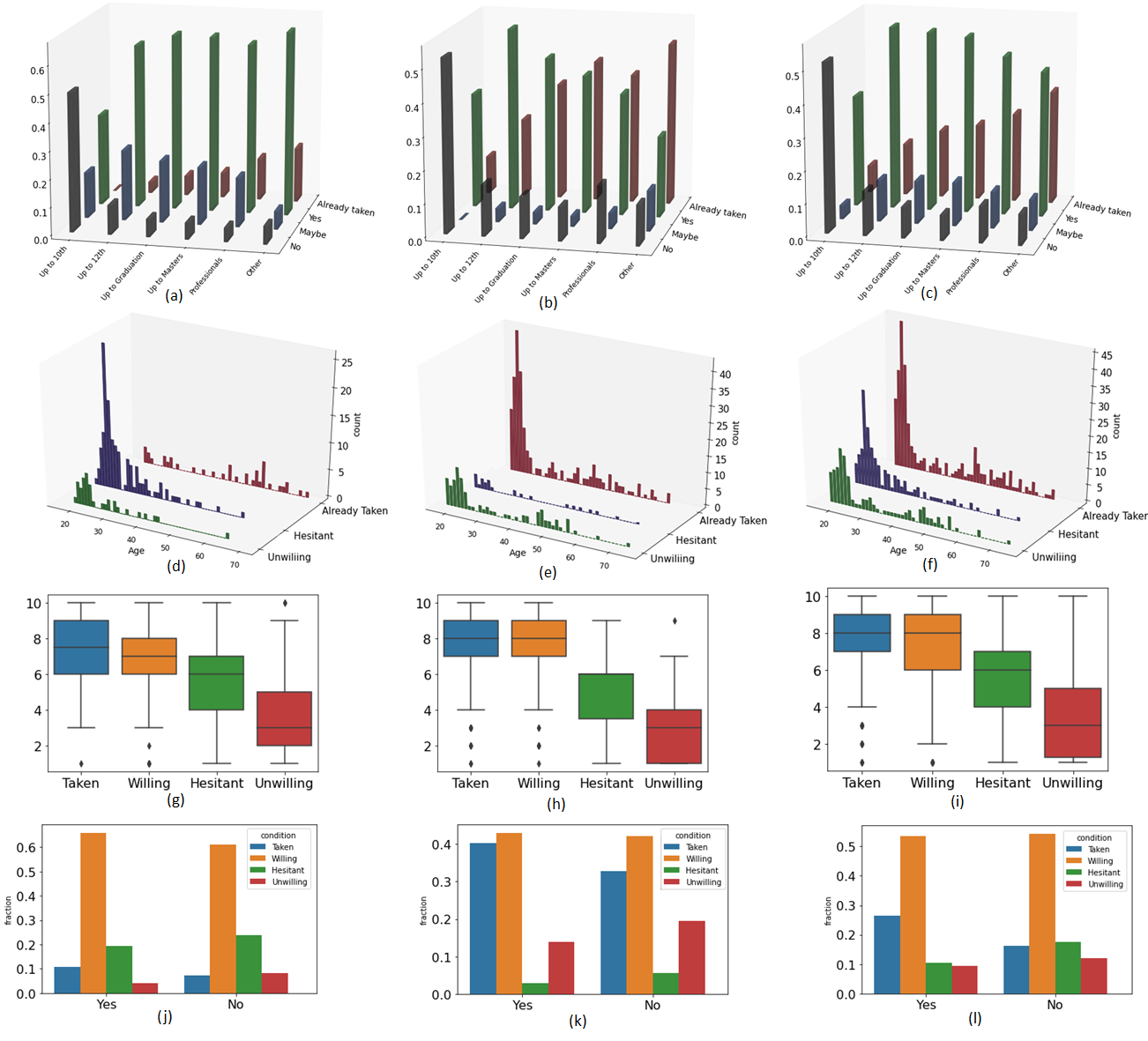}
\caption{\label{fig:fig2}   Normalized distribution of vaccine willingness depending on educational background (a) in the $1^{st}$ phase; (b) in the $2^{nd}$ phase; (c) on the entire data.  Vaccine willingness depending on age (d) in the $1^{st}$ phase; (e) in the $2^{nd}$ phase; (f) on the entire data. The perception about the effectiveness of the vaccine among different subpopulations (g) in the $1^{st}$ phase, (h) $2^{nd}$ phase; (i) on the entire data. The effect of hyperbolic discounting (j) in the $1^{st}$ phase, (k) $2^{nd}$ phase; (l) on the entire data. Here, the $1^{st}$ phase indicates the time from the beginning of the survey to the date when the peak of the second infection wave occurs, whereas the $2^{nd}$ phase indicates the time between the occurrence of the second wave peak and the survey end-date. In case of hyperbolic discounting analyses, `Yes' indicates the group of people who exhibited hyperbolic discounting, and `No' indicates the rest of the people. }
\end{figure}
\section{Results}
%\subsection{Data Understanding in Multi-feature Space}

Details regarding the conducted survey are discussed in the supplementary material. Contingent on the degree of willingness that respondents display to get vaccinated, they are asked an additional set of questions which are different for each group. Two rounds of the same survey described above are administered before and after the peak of COVID-19 second wave in India in 2021. The geographical locations extracted from the PIN codes indicated by the respondents show that almost the entire country has been covered (Fig. \ref{fig:fig1}(a) ). Some of the key statistical findings related to age, gender, family income, education COVID exposure etc. are shown in Fig. \ref{fig:fig1}(b). Depending on the willingness to take the vaccine, the entire respondent pool has been divided into four groups$-$ willing people (`W'), who want to take the vaccine if it is available; hesitant people (`H'), who are not sure about taking the vaccine; unwilling people (`U'), who do not want to take the vaccine, and finally the people who have already taken at least one dose of the vaccine (`V'). \\

There are several contributing factors in the decision to get vaccinated. We identify that compared to individuals with a higher level of education, in the group having educational qualifications only up to 10$^{th}$ standard, the fraction of people who are unwilling to take the vaccine is relatively high. We also identified that before the occurrence of the peak of the second wave, younger people were more unwilling to take the vaccine. However, these trend had changed significantly once the peak of the second wave occurred. Similar interesting trends were observed for vaccine effectiveness perception and behavioural factors like hyperbolic discounting as shown in in Figs. \ref{fig:fig2}(a)-(l). However, it is challenging to quantify the impact of all the contributing factors due to their complex inter-dependencies.Thus, to understand the marginal effects of each survey variable on vaccine hesitancy, we choose an econometric analysis framework, and proceed with a comprehensive Logit estimation \cite{wooldridge2010econometric} of our data.  
\subsection{Binary Logit Estimation Results}

\begin{table}
\caption{Marginal Effects from Binary Logit}
{\scriptsize{
\begin{tabular}{|l|c|c|c|} 
 \hline
 Variables & Vaccine Hesitancy (1) & Vaccine Hesitancy (2) & Vaccine Hesitancy (3) \\ [0.1ex] 
 \hline
 Self-Contract COVID & -0.079***	& -0.067**	 & -0.066**  \\ 
& (0.028) &	(0.028)&	(0.027)\\
 COVID Contracted among Friends/Family & 0.017 & 0.018 & 0.007 \\
 & (0.025) & (0.026)& (0.026) \\
 COVID Deaths among Friends/Family &	-0.062*** &	-0.049** &	-0.048** \\
	&(0.021)&	(0.022)&	(0.022) \\
Vaccine Effectiveness Perception &	-0.075*** &	-0.075*** &	-0.074*** \\
&	(0.003)&	(0.003)	& (0.003) \\

Risk Aversion&	0.013&	0.017&	0.023 \\
&	(0.026)&	(0.026)&	(0.026) \\
Loss Aversion&	-0.026&-0.028&	-0.029 \\
	&(0.025)&	(0.025)&	(0.025) \\
Hyperbolic Discounting&	-0.085***&	-0.085***&	-0.096***\\
	&(0.026)	&(0.025)&	(0.026) \\
Impatience&	-0.012&	-0.017&	-0.015 \\
	&(0.028) &	(0.028)&	(0.028) \\
Age	&-0.003**&	-0.002&	-0.002 \\
&(0.001)	&(0.001)&	(0.001) \\
Female	&-0.037*&	-0.033&	-0.037*\\
	&(0.021)&	(0.021)&	(0.022)\\
Education (Base category: Up to School Std. 10)&	$--$&	$--$  &	$--$\\
Education - School Std.12	&-0.240**&	-0.275*** &	-0.261***\\
	&(0.101)&	(0.104)&	(0.099)\\
Education- Graduation	&-0.248**&	-0.290***&	-0.290***\\
	&(0.096)&	(0.099)&	(0.095)\\
Education- Post-Graduation	&-0.242**&	-0.292***&	-0.280***\\
&	(0.097)&	(0.099)&	(0.095)\\
Education- Professional Degree	&-0.269***&	-0.306***&	-0.314***\\
&	(0.100)&	(0.104)&	(0.100)\\
Education- Other&	-0.196*	&-0.244*&	-0.271**\\
&	(0.118)&	(0.126)&	(0.125)\\
Occupation (Base category-Student)&	$--$  &	$--$  &	$--$\\
Occupation-Govt. organization employee&	0.001&	0.005&	-0.012\\
&	(0.042)&	(0.043)&	(0.041)\\
Occupation-Private organization employee&	0.079**&	0.060*&	0.057*\\
	&(0.035)&	(0.035)	&(0.035)\\
Occupation- Self-employed	&0.138**&	0.116*&	0.092\\
	&(0.062)&	(0.06)	&(0.058)\\
Occupation- Retired	&0.019	&-0.029	&-0.031\\
&	(0.121)	&(0.112)&	(0.116)\\
Occupation- Housewife&	0.108&	0.092 &	0.115\\
&	(0.078)	&(0.076)&	(0.076)\\
Occupation -Not Employed&	0.147&	0.134	&0.060\\
	&(0.104)&	(0.099)&	(0.095)\\
Occupation- Other&	0.049	&0.032	&0.020\\
&	(0.057)	&(0.057)&	(0.061)\\
Location&	-0.00001&	-0.00003&	9.17e-06\\
&(0.00012)&	(0.00012)&	(0.0001)\\
Surveyed Post COVID-19 Second Wave Peak&	$--$  &	-0.08***&	-0.076***\\
	&&	(0.022)	&(0.022)\\
Family Income (Base category- Less than 5 lacs)&	$--$ &	$--$  &	$--$\\
Family Income- Between 5-10 lacs &	$--$ &	$--$ & 0.027\\
			& & & (0.027) \\
Family Income- Between 10-20 lacs & $--$ & $--$ & 0.016 \\
&-&-& (0.028)\\
Family Income- Between 20-30 lacs & $--$ & $--$ & 0.036 \\
&&& (0.040)\\
Family Income- Above 30 lacs &&& 0.065 \\
&&& (0.040)\\
 \hline
Observations &	1193 &	1193 &	1141 \\
Pseudo R2 &	0.298 &	0.308 &	0.321 \\
\hline
\hline
\end{tabular}
}
\\ Robust Standard errors in parentheses
 *** $p < 0.01$, ** $p < 0.05$, * $p < 0.1$}
\end{table}

We estimate the impact of our variables of interest using a logistic regression. For  the  purpose  of our estimation,  We define our dependent 'vaccine hesitancy' as a binary variable that takes the value 1 if an individual’s response is ‘maybe’ or ‘no’ to the question on the survey that asks, ‘do you wish to get vaccinated?’ and 0 otherwise. The details of the analysis are mentioned in Materials \& Methods. We run three versions of the estimating equation (check eq. 1 in supplementary). The first version includes Covid-19 related variables, behavioural variables and some basic socio-demographic variables. The second version includes the first version plus a variable that indicates whether the observations are from before or after the peak infection levels for the second wave of COVID-19 in India. Finally, version 3 considers the full list of $k$ variables and includes the second version as well as the income bands to which the individuals belong.\\ 

Logistic regression coefficients report changes in log-odds for the dependent with a unit change in the independent variable. However, for purposes of interpretation, marginal effects that report the change in likelihood for the dependent variable for a unit change in a regressor are more useful. Accordingly, Table 1 reports marginal effects from logit estimation models for the outcome variable vaccine hesitancy among adults in India. Columns 1, 2 and 3 present results from the three versions of our logistic regression model we describe above. We find that across our three models, contracting COVID lower an individual's vaccine hesitancy between 6.6 and 7.9 percentage points (pp) while COVID-19 deaths in one's family and social network reduces the same between 4.8 and 6.2 pp. These effects are significant at least at the $p<0.05$ level in all cases. Our second important result is that a unit increase in perceived effectiveness of the COVID-19 vaccine is associated with about 7.5 pp lower hesitancy. This effect is highly significant at the $p<0.01$ level in all three models.\\

Similar to Barello et. al. [1], we investigate the impact of behavioral and psychological factors on individuals’
attitudes towards vaccines. We find that individuals exhibiting quasi-hyperbolic discounting in their time preference are between 8.5 and 9.6 pp less likely to be vaccine hesitant and this is significant at $p < 0.01$ level. An individual exhibiting quasi-hyperbolic discounting of time preference (sometimes also known as 'present bias') is defined as somebody who chooses the option of getting the lower amount right now (question: if you had a choice between getting Rs. 2000 right now versus Rs.4000 in six months’time, what would you choose) but is ready to wait for the higher amount in the future (question: if you had a choice between getting Rs.2000 in twelve months versus getting Rs. 4000 in a year and six months, what would you choose?) and thereby exhibiting dynamic inconsistency in time preference. We find that other behavioral factors such as impatience, risk and loss aversion are not significant predictors of vaccine hesitancy.\\

Among socio-demographic factors, we find that age, gender and education are significant determinants of vaccine hesitancy. Specifically, for our first model, we find that with 1 year increase in age, individuals’ likelihood of vaccine hesitancy reduces by 0.3 pp. We also find that women are 3.7 pp less likely to exhibit vaccine hesitancy and it is weakly significant at $p<0.1$ level. The effects for education are stronger and more significant. In our regression models. individuals having 10$^{th}$ Standard as their highest educational qualification are considered as the base category for the level of education of an individual. As compared to this base category, we find that for individuals having 12$^{th}$ Standard as the highest level of education, likelihood of vaccine hesitancy reduces by 24 - 27.5 pp ($p<0.05$). For graduates, as compared to the base category, the likelihood of vaccine hesitancy reduces by 24.8 - 29 pp ($p<0.05$) while for postgraduates and individuals with professional degrees it reduces by 24.2-29.2 pp ($p<0.05$) and 26.9-31.4 pp ($p<0.05$) respectively over the three models. Our findings are consistent with \cite{robertson2021predictors} who observe that hesitancy levels are higher for individuals with lower levels of education. We also find that as compared to the base category of students, individuals employed in private organizations are 7.9 pp and self-employed individuals are 13.8 pp ($p<0.05$) more likely to be vaccine hesitant in model 1. %The private organization indicator is only marginally significant ($p<0.1$) in column 2 and 3, while the self employed indicator is marginally significant ($p<0.1$) in column 2 and insignificant in column 3.%  
One important finding is that individuals who were surveyed after the peak of second wave are between 7.6-8 pp less likely to be vaccine hesitant and this result is highly significant at $p<0.01$ level. Finally, from Column 3 we find that controlling for family income does not change any of our significant findings from columns 1 and 2.  \\
 
As we collect data over four ordered categories for our hesitation variable in our survey, we also run as robustness check, ordered logit estimations which are straightforward generalizations of the binary framework above \cite{wooldridge2010econometric}. The results are robust with those from our binary logit estimates and can be found in the supplementary material. \subsection{Understanding Population-level Impact: Network Simulations}
\begin{figure}
\centering
\includegraphics[width=\textwidth]{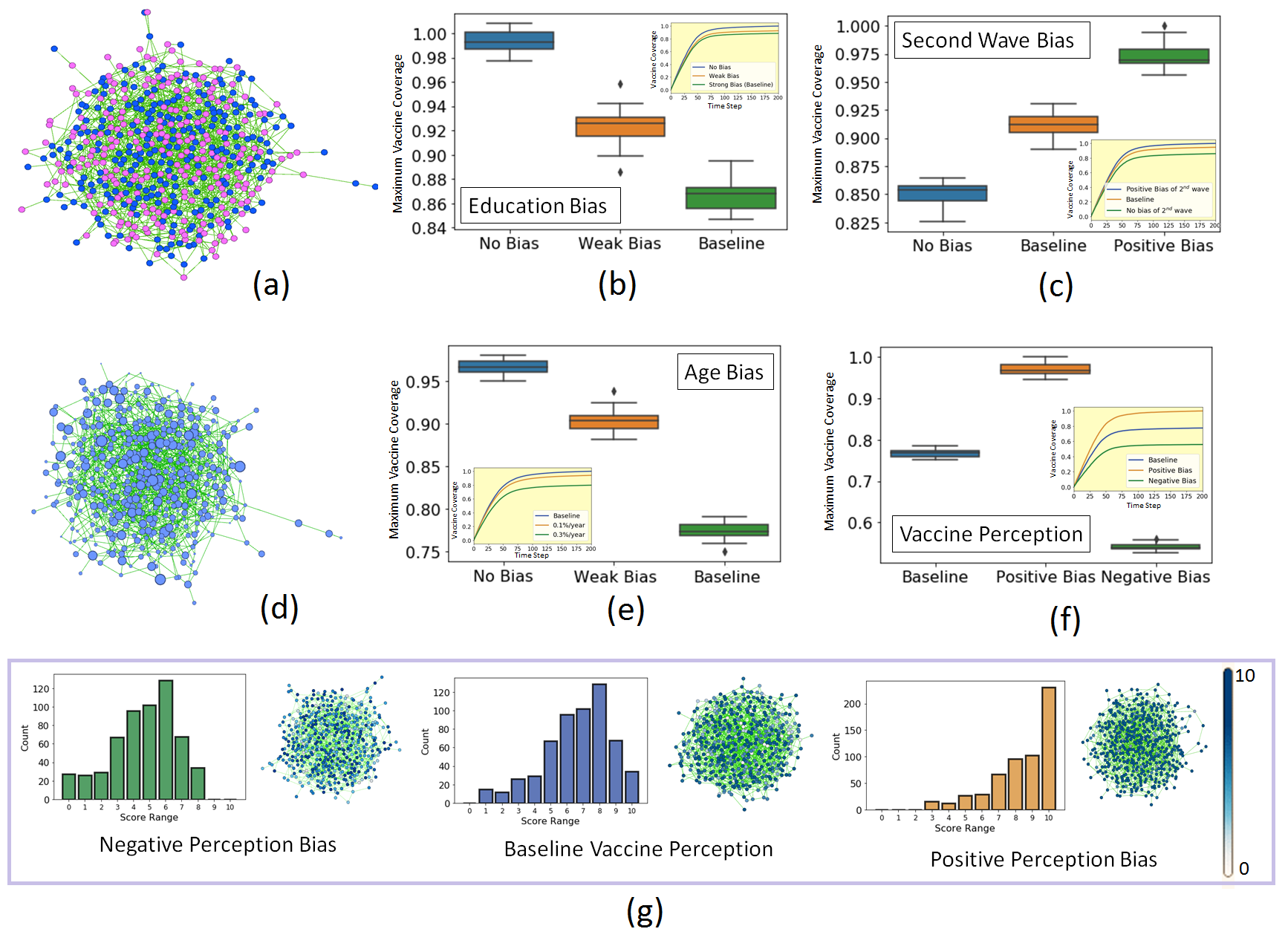}
\caption{Effect of major barriers: (a) and (b) shows the effect of education bias. (a) The network architecture with education bias attribute assigned to each node. (b) Box plot of maximum vaccination achieved on the artificial society. (c) Effect of encountering second-wave in form box plots. (d) and (e) shows the effects of age-driven factor on vaccine coverage. (d) The network architecture with age bias attribute assigned to each node. (e) Box plot of maximum vaccination achieved on the artificial society. (f) Effects of positive and negative vaccine efficiency perceptions compared to the baseline perception on vaccine coverage (g) Original and shifted histograms, along with network architecture for vaccine efficiency perception score, used for simulations shown in (f). (Insets of (b), (c), (e) and (f) shows the time evolution of vaccination coverage in the synthetic societies)  
}\label{fig:fig3}
\end{figure}
From our binary logit estimation analysis (as shown in Table 1), we detect some major drivers and barriers of the dynamics: (i) perception of vaccine effectiveness, (ii) hesitancy in younger population and (iii) hesitancy due to lack of education. Unlike other factors, these three statistics about an individual are easily available for policy making and can steer the vaccination dynamics. To establish the impact of these factors we have identified, we consider a synthetic heterogeneous population and impose different constraints to observe the variations of vaccine coverage under such conditions.\\

To achieve this, we use compartmental models of infectious disease propagation on complex networks \cite{epidem, bauch2005dynamically, ghosh2020data}. We take a similar model with SEIQRV (Susceptible-Exposed-Infected-Quarantined-Recovered-Vaccinated) structure on a complex random network (\textit{see} Materials \& Methods), that takes into account of the vaccination along with disease spreading. \\

As pointed out in Table 1, education plays an important role in vaccine decision making. Compared to the base category, i.e., people who have educational qualification up to 10$^{th}$ standard, people with higher educational qualification are significantly less vaccine hesitant. To understand the impact of education on the vaccination coverage of the society, we split our synthetic population in two parts: $-$ one having educational qualification up to 10$^{th}$ standard (pink nodes in Fig. \ref{fig:fig3} (a)) and others having higher educational background (blue nodes in Fig. \ref{fig:fig3} (a)).\footnote{To decide the fraction of pink nodes, we take into account the current ratio of Indian population with below and above 10$^{th}$ standard education} As per Table 1, the pink nodes are 26 pp more likely to be hesitant to vaccination and this is taken as the baseline scenario. Considering two more situations where the pink nodes are 13 pp more hesitant (weak bias) or have equal vaccine acceptance as blue nodes (no bias), we see that (Fig. \ref{fig:fig3}(b)), reduction of vaccine hesitancy among the pink nodes can increase vaccine coverage by 15.6 pp, which is a substantial improvement.\\

Now, we focus on the second wave bias and study the population level impact of the logit analysis result. To understand the impact, we evaluate the maximum vaccine coverage under three different situations: no bias (no change in vaccine hesitancy post-second-wave), baseline (people are 7.6 pp less hesitant after second wave) and finally, positive bias (people are 15 pp less hesitant post-second-wave). As shown in Fig. \ref{fig:fig3} (c), sensitivity towards the impacts of second wave causes a considerable ~14.7 pp more vaccination coverage in post-second-wave societies. Apart from education, logit analysis points out age to be a major reason of hesitancy. %with a rate of 0.3 pp/year decrease in hesitancy. 
%For a country like India where the average age of population is much on the lower side, this might be one of the major reasons behind the reduced vaccination coverage. To quantify this effect, we proceed to study the dynamics on a simulated society where either this age-dependant hesitancy is absent or it is present in a lower extent (say, 0.1\%/year). 
The network structure embedded with the current age distribution of India is shown in Fig. \ref{fig:fig3}(d) where node size depicts the age of that individual. We report three circumstances: baseline (as per logit results, 0.3 pp/year), weak bias (hesitancy in lower extent, 0.1 pp/year), and no bias (no age-dependent hesitancy. The results are shown in Fig. \ref{fig:fig3}(e) clearly shows that the present hesitancy in the younger population in India is causing 19.9 pp less vaccine coverage in our synthetic population compared to the maximum vaccine coverage possible.\\

Finally, we study the dynamics with the perceived effectiveness of the vaccine as another attribute. In agreement with our findings reported in Fig. \ref{fig:fig2}, people with higher value perceived efficiency score are less likely to be vaccine hesitant. To implement that into our synthetic society, we define perception score $p_s$ as a node attribute. Next, introducing a moderate improvement or deterioration of this perception score ($p_s \pm 2$), we study three possible scenarios. In Fig.\ref{fig:fig3}(g), the distributions for baseline vaccine perception (extracted from survey data) and the variations considered are shown along with the networks with modified vaccine perceptions (different shades of blue color depict different levels of vaccine efficiency perception). 
%The positive (negative) perception bias increases (decreases) the nodes possibility to take the vaccine compared to the baseline case. 
Results depict (Fig. \ref{fig:fig3}(f)) that compared to the existing scenario (baseline) a positive (negative) bias can improve (degrade) the maximum vaccine coverage up to 20 pp in our experiments.  Considering that $p_s$ is just a perceived score (not clinically reported vaccine efficiency), and could be improved by awareness campaigns and promotions, this result has an important strategic and policy implication.

\begin{figure}
\centering
\includegraphics[width=\textwidth]{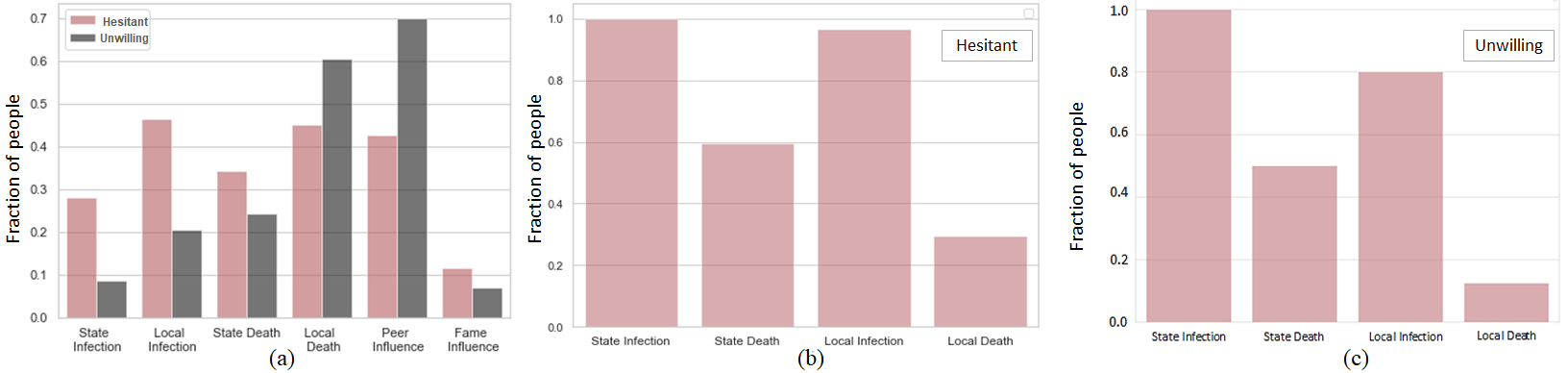}
\caption{\label{fig:fig4} (a) Different factors that affect the vaccination decisions of unwilling and hesitant sub-populations; (b) \& (c) Erroneous perceptions of reality for hesitant and unwilling sub-populations (refer to the text for more details). }
\end{figure}
\subsection{\textcolor{black}{Decisive Factors, Information Gap \& Reality Check}} 
\textcolor{black}{Before we proceed with the prescription of some strategic interventions, we point out another important fact associated with our results. While the responses gathered through the survey, reflects people's perceptions, opinions and their understanding of epidemic and vaccination scenario, it is very important to note that there might exist some gap between the reality and these perceptions. For example, from responses given by the hesitant and unwilling people, we analyzed the factors that drive to change their vaccination decision. These results are consolidated in Fig. \ref{fig:fig4}(a). We found that people from `H' sub-population are willing to take the vaccine if the infection level or the death toll rises in the state. However, if the local death cases rise, unwilling people (U) are more willing to take the vaccine compared to hesitant sub-population. In our analyses, we find that for both `H' and `U' sub-populations, peer influence is a strong factor and they are willing to take the vaccine if someone from their family or close friends take the vaccine. Compared to the news of any celebrity taking the vaccine, influence of peers have much more effective impact for both the sub-populations. In fact, peer influence turns out to be the strongest factor to change the decision of unwilling people compared to the other pandemic related parameters, like state-level infection, local infection, state-level death counts or local death counts.}\\

Based on the above findings, we proceed with a reality check analysis. As we have seen from the data that local rise of infection and death toll have shown strong implications on the vaccine hesitancy, we further investigate how the actual spread of the pandemic influences the decisions of vaccine hesitant people. Interestingly, we figured out that these factors depend a lot on how informed the people are about the actual scenario. As we had recorded the PIN codes and the answer submission time-stamps of all the respondents, we tallied their decisions with actual COVID-19 statistics from their geographic locations.  For hesitant or unwilling particular sub-population, if a respondent says that the he/she will be willing to take the vaccine if the state(local) infection(death) increases, we check whether the infection(death) at that state(district) was increasing or not for last 10 days from the time of submitting his/her response. If the infection(death) was increasing, we can conclude that the respondent was unaware about the actual situation of COVID-19 and thus, making the vaccine decision which was not coherent with the reality as well as their response. While details of the methods are mentioned in Materials \& Methods, we report how the fraction, $\mathcal{R}$ stands for both $H$ and $U$ individuals in Figs. \ref{fig:fig4}(b) and (c). In these figures, the y-axes indicate the fraction of people, $\mathcal{R}$ with a misconception about the real COVID scenario. Our results imply that for a considerable percentage of people, their decisions are based on perceptions far from reality. They are thinking of taking the vaccine if the state(local) infection(death) rises while they are unaware about that rise already happening at that time. As shown in Figs. \ref{fig:fig4}(b) and (c)  for both the `H' and `U' sub-populations, almost all the people who were vaccine hesitant, made their decisions contradicting the reality based on state (local) infection rise. This analysis provides an important observation that many people are not dropping their hesitations and unwillingness just because of the lack of information. Thus, proper dissemination of accurate local level COVID-19 situation might increase the vaccine coverage significantly.

\section{Discussion \& Concluding remarks}
%\textcolor{blue}{- Cover almost entire country\\
%- Important factors from logit analyses\\
%- Simulation on synthetic society to detect how the number of %vaccinated people can change depending on the factors we have identified\\
%- Importance of peer influence (which we will be modeling in paper 2)\\
%- Reality check using spatial information\\
%- Omicron\\
%- Some policies\\}

From our extensive primary survey, we model vaccine hesitancy to be a multidimensional socio-economic phenomenon. We use a logit framework to identify the main variables that are significantly associated with vaccine hesitancy. These key drivers can be categorized in four contextual clusters. First, issues related to uncertainty regarding reliability and effectiveness of the available vaccines. Second, individual-specific issues, associated with gender, younger age or lack of education. Third, infection-specific issues that relate to encountering an epidemic wave and observing its consequences. Finally, informational factors related to exposure to COVID-19 awareness campaigns and local level peer influence. \\

We take our study one step further and simulate epidemic dynamics with spontaneous vaccination on a synthetic heterogeneous society to measure the consequences caused by these factors on population-level vaccine coverage. We identify quantified estimates for improved (or hindered) vaccination coverage, considering inputs that are statistically significant from our logit estimation results. The inputs used for our simulations are age, perceived vaccine effectiveness, experience of a COVID-19 wave and education. \textcolor{black}{Our simulation results show that encountering the devastation of a COVID-19 wave can lower the hesitancy of a society as a whole, and improve the number vaccinated people by a considerable percentage. The predictive model also shows that age bias, education bias and vaccine effectiveness perception bias are capable of causing a substantial 15-20 pp reduction in the number of people getting vaccinated.} Finally, we estimate the importance of the information gap in the opinions expressed by the individuals in our survey, using an exhaustive data mining analysis. Employing submitted PIN codes by respondents, we verify that most of our hesitant and unwilling sample have incorrect beliefs regarding the local level spread of infection. These incorrect beliefs impact their willingness to be vaccinated and in certain cases promote hesitancy.Thus the central and state governments have to step up local level awareness campaigns which will convey to residents the level of seriousness of infections and hospitalizations in their localities. As the young and less educated seem much more hesitant towards getting vaccinated, targeted information campaigns have to be implemented to push behaviour change among these demographics. Finally, our results underscore the importance of vaccine effectiveness perception on expressed vaccine readiness. Therefore, administrations need to continue to stress not just the importance of being double (or triple) vaccinated, but of providing accurate information about their efficacy.

\section*{Author contributions:}
Conceptualization: SG, SB, SM, SC\\
Methodology: SM, SB, SC, SG\\
Investigation: SM, SB\\
Visualization: SB, SG\\
Writing: SC, SG, SB, SM

\bibliographystyle{abbrv}
\bibliography{scibib}

\end{document}